\documentclass[preprint,aps,nofootinbib,tightenlines]{revtex4}

\usepackage{graphicx}
\usepackage{amsmath}

\newcommand{\babar}{\mbox{\ensuremath{{\displaystyle B}\!{\scriptstyle A}
{\displaystyle B}\!{\scriptstyle AR}}}}
\newcommand{\BtoXulnu}{\ensuremath{B\rightarrow X_u \ell\nu}}
\newcommand{\BtoXclnu}{\ensuremath{B\rightarrow X_c \ell\nu}}

\newcommand{\BtoXsgamma}{\ensuremath{B\rightarrow X_s \gamma}}

\newcommand{\BtoXgamma}{\ensuremath{B\rightarrow X \gamma}}
\newcommand{\Vub}{\ensuremath{|V_{ub}|}}
\newcommand{\Vcb}{\ensuremath{|V_{cb}|}}

\newcommand{\Eg}{\ensuremath{E_\gamma}}
\newcommand{\Emin}{\ensuremath{E_\gamma^\mathrm{min}}}

\newcommand{\mxsq}{\ensuremath{m_X^2}}
\newcommand{\mes}{\ensuremath{m_\mathrm{ES}}}
\newcommand{\mb}{\ensuremath{m_b}}
\newcommand{\mupisq}{\ensuremath{\mu_\pi^2}}
\newcommand{\muGsq}{\ensuremath{\mu_G^2}}
\newcommand{\rhoDcube}{\ensuremath{\rho_D^3}}
\newcommand{\rhoLScube}{\ensuremath{\rho_{LS}^3}}
\newcommand{\Breco}{\ensuremath{B_\mathrm{reco}}}
\newcommand{\mommk}{\ensuremath{\langle m_X^k \rangle}}
\newcommand{\momnk}{\ensuremath{\langle n_X^k \rangle}}
\newcommand{\unit}[1]{\,\mathrm{#1}}
\begin{document}
\preprint{\vbox{ \hbox{SLAC-PUB-13036} } }

\title{\boldmath
Determination of the $b$-quark mass and nonperturbative parameters
in semileptonic and radiative penguin decays at \babar}

\author{Kerstin Tackmann (on behalf of the \babar\ collaboration)}

\affiliation{Ernest Orlando Lawrence Berkeley National Laboratory,
University of California, Berkeley, CA 94720
\vspace*{3ex}}

\begin{abstract}
Knowing the mass of the $b$-quark is essential to the
study of the structure and decays of $B$ mesons as well
as to future tests of the Higgs mechanism of mass generation.
We present recent preliminary measurements of the $b$-quark 
mass and related nonperturbative
parameters from moments of kinematic distributions in charmed and charmless 
semileptonic and  
radiative penguin $B$ decays. Their determination from charmless 
semileptonic $B$ decays is the first measurement in this mode.
The data were collected by the \babar\ detector at the PEP-II asymmetric-energy
$e^+e^-$--collider at the Stanford Linear Accelerator Center at a 
center-of-momentum
energy of $10.58\unit{GeV}$.
\end{abstract}
\maketitle
\section{Introduction}
An important goal of the $B$-physics program is the precise measurement 
of the CKM matrix~\cite{Cabibbo:1963yz}
elements \Vcb\ and \Vub. The most accurate determinations are
obtained from semileptonic decays \BtoXclnu\ and \BtoXulnu, respectively.
Generally, two different approaches can be used: The hadronic state 
$X_{c,u}$ can be reconstructed either in specific exclusive modes, 
or inclusively by summing over all possible hadronic final states.
The inclusive determinations rely
on an Operator Product Expansion (OPE) in inverse powers of the $b$-quark mass
\mb~\cite{Bigi:1993fe}. 
At second order in the expansion, two nonperturbative parameters arise, 
which describe the kinetic energy and the chromomagnetic moment of the $b$ 
quark
inside the $B$ meson. In the kinetic scheme, they are denoted by \mupisq\ and
\muGsq, respectively. Two more parameters arise at third order, 
\rhoDcube\ and \rhoLScube. In the kinetic scheme, short- and long-distance
contributions are separated by a hard cutoff $\mu$ and the $b$-quark mass and
nonperturbative parameters are given at $\mu=1\unit{GeV}$.

The mass \mb\ and the nonperturbative parameters can be determined from the 
study of
kinematic distributions in semileptonic and radiative penguin $B$ decays.

Precise measurements of \mb\ are needed both to reduce the uncertainty of
inclusive determinations of \Vub, as well as for studying New Physics effects
in the Higgs sector at future experiments.

Here, we present recent determinations of \mb\ and higher-order nonperturbative
parameters from charm and charmless semileptonic and radiative penguin $B$ 
decays at the \babar\ experiment~\cite{Aubert:2001tu}.
\section{The Recoil Method}
In all analyses presented, $\Upsilon(4S) \to B\bar{B}$ decays are tagged
by reconstructing one $B$ meson (\Breco ) fully in hadronic modes, 
$\Breco\to D^{(*)} Y^\pm$. The $Y^\pm$ system is composed of hadrons with a
total charge of $\pm 1$, 
$Y^\pm = n_1 \pi^\pm + n_2 K^\pm + n_3 K_S + n_4 \pi^0$, with
$n_1 + n_2 \leq 5$, $n_3 \leq 2$, $n_4 \leq 2$. We test the kinematic
consistency of \Breco\ candidates with two variables, 
$\mes = \sqrt{s/4 - \vec{p}_B^{\,2}}$
and $\Delta E = E_B - \sqrt{s}/2.$ Here, $\sqrt{s}$ is the invariant mass
of the $e^+e^-$ system and $E_B$ and $\vec{p}_B$ denote the energy and momentum
of the \Breco\ candidate in the $\Upsilon(4S)$ frame. We require $\Delta E$ to
be $0$ within three standard deviations.
In events with multiple \Breco\ candidates we retain the candidate 
reconstructed in the mode with 
highest purity as estimated from the ratio of signal over background for
events with $\mes > 5.27\unit{GeV}$ on Monte Carlo simulation 
(MC).

By fully reconstructing one of the $B$ mesons in the event, the charge,
flavor and momentum of the second $B$ can be inferred. All particles that
are not used in the reconstruction of the \Breco\ are assigned to
the decay of the signal $B$.
The efficiency to reconstruct a \Breco\ candidate is $0.3\%$ ($0.5\%$) for
$B^0\bar{B}^0$ ($B^+B^-$) events.
\begin{figure}[t]
    \begin{center}
        {\includegraphics[width=1.\textwidth]{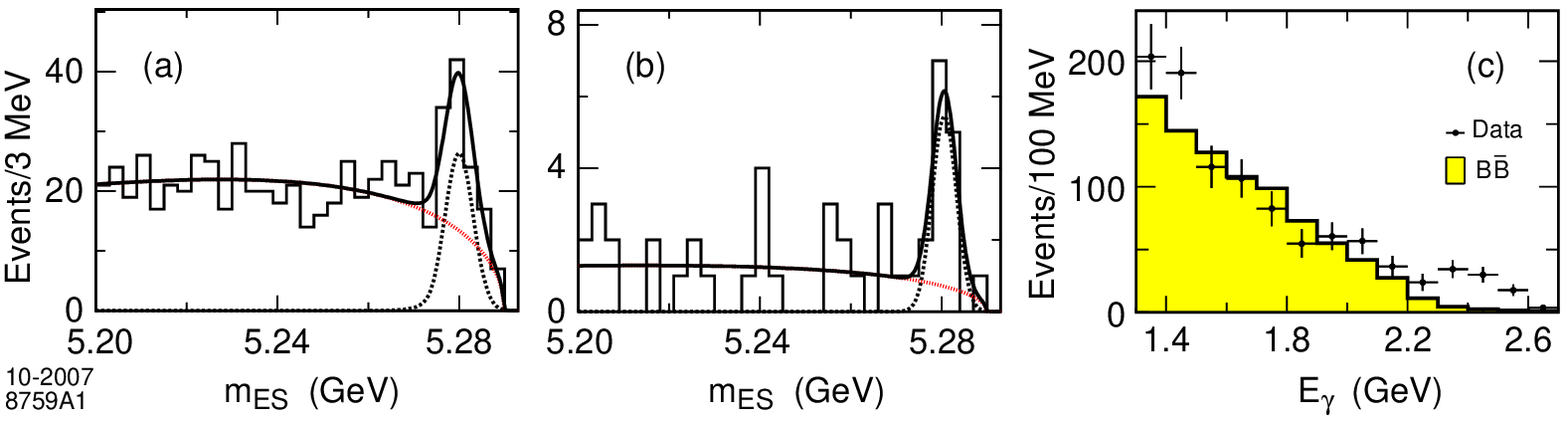}}%
        \caption{\it Fits in \BtoXsgamma\ to \mes\ for two \Eg\ regions. 
The dashed curve shows
the signal, the solid grey (red) curve the background as given by the fit. 
The solid black curve
shows the sum of signal and background. 
(a) $1.6\unit{GeV} < \Eg < 1.7\unit{GeV}$ for charged $B$ events and
(b) $2.3\unit{GeV} < \Eg < 2.4\unit{GeV}$ for neutral $B$ events. 
(c) The measured photon energy spectrum before subtraction of backgrounds.
Points show the data spectrum, the shaded histogram shows the $B\bar{B}$ 
backgrounds, where the shape is taken from MC.}
\label{btosfig}
    \end{center}
\end{figure}

We use fits to the \mes\ distribution to subtract the combinatorial background
from $B\bar{B}$ events and and the background from continuum 
($e^+e^- \to q\bar{q}, q=u,d,s,c$) events.
The backgrounds are modeled with a threshold function~\cite{Albrecht:1990cs}
and the signal is
described by a Gaussian function joined with an exponential tail to describe
photon energy loss~\cite{Skwarnicki:1986xj}. Examples from the measurement in
radiative penguin decays are shown in Fig.~\ref{btosfig} (a) and (b).
\section{Radiative Penguin Decays}
The first and second moments of the photon energy spectrum in radiative 
penguin 
decays, \BtoXsgamma, are used for a preliminary determination of \mb\ and 
\mupisq~\cite{bsg}. The moments are
extracted as a function of the lower cut on the photon energy, \Emin,
measured in the rest frame of the signal $B$. 
The full reconstruction of the second $B$ in the event  results in an improved
signal purity and different systematic uncertainties, but lower statistics, 
than alternative methods.

The measurement is based on a sample of $232$ million $B\bar{B}$ pairs.
Events with a well reconstructed, high energy photon are selected if 
the photon is not compatible with originating from the decay of a $\pi^0$ or 
$\eta$, or a $\rho^\pm \to \pi^\pm \pi^0$ decay assuming that the second 
photon from
the $\pi^0$ decay was lost. Continuum background is
suppressed by using a Fisher discriminant that makes use of the difference 
between event topologies in $B\bar{B}$ and continuum events.

The \Eg\ spectrum is measured in bins of $100\unit{MeV}$ and is shown in
Fig.~\ref{btosfig} (c). The region 
$1.3\unit{GeV} < \Eg < 1.9\unit{GeV}$ is used to normalize the backgrounds, 
the largest part of which consists of photons from unreconstructed $\pi^0$ 
or $\eta$ decays.
The backgrounds are extrapolated into the signal region, 
$\Eg > 1.9\unit{GeV}$, the shape of the backgrounds is taken from MC. 
The signal region contains $119\pm 22$
\BtoXgamma\ events over an estimated background of $145\pm 9$ events.
The measured photon spectrum is corrected
for efficiency, which varies with \Eg, and resolution effects.
First and second central moments, $\langle \Eg \rangle$ and 
$\langle \Eg^2 \rangle - \langle \Eg \rangle^2$, as a function of \Emin\ 
are extracted from the corrected spectrum. The moments 
for $\Emin \leq 2.0\unit{GeV}$
are used to determine $\mb = 4.46^{+0.21}_{-0.23}\unit{GeV}$ and 
$\mupisq = 0.64^{+0.39}_{-0.38}\unit{GeV}^2$ in the kinetic scheme with a 
correlation of $\rho = -0.94$.
\section{Charm Semileptonic $B$ Decays}
Moments of the hadronic mass and lepton energy spectra in \BtoXclnu\ decays 
are used for a preliminary extraction of
\mb, the charm-quark mass $m_c$, nonperturbative parameters and 
\Vcb~\cite{:2007yaa}. We present moments of 
the hadronic mass \mommk, $k=1..6$, which use a larger data set than the 
previous 
measurement and new measurements of the mixed hadronic mass-energy moments 
\momnk, $k=2,4,6$. 
$n_X$ is defined by 
$n_X^2 = m_X^2 c^4 - 2\bar{\Lambda} E_X + \bar{\Lambda}^2$, where $m_X$ is
the mass and $E_X$ the energy of the inclusive $X_c$ system in the $B$ rest
frame and $\bar{\Lambda}=0.65\unit{GeV}$.
The mixed moments are expected to yield a more precise determination 
of higher order
nonperturbative parameters. The moments are extracted as a function of
a lower cut on the lepton energy between $0.8\unit{GeV}$ and $1.9\unit{GeV}$
in the signal $B$ rest frame.

\begin{figure}[t]
    \begin{center}
        {\includegraphics[width=0.532\textwidth]{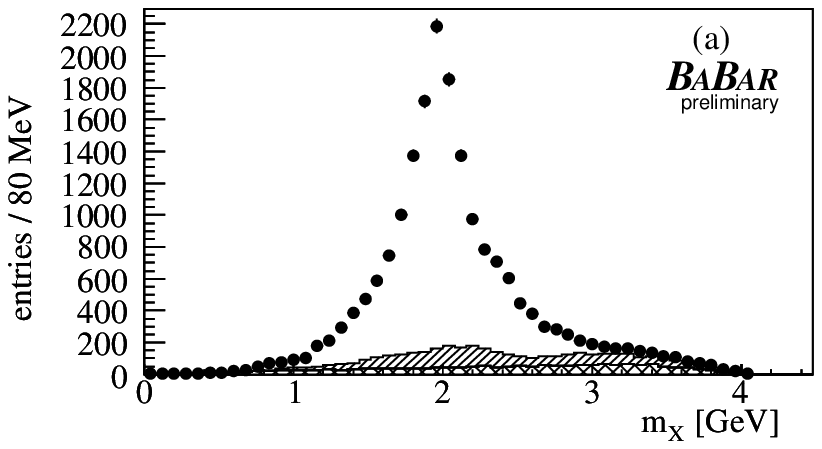}}%
        {\includegraphics[width=0.468\textwidth]{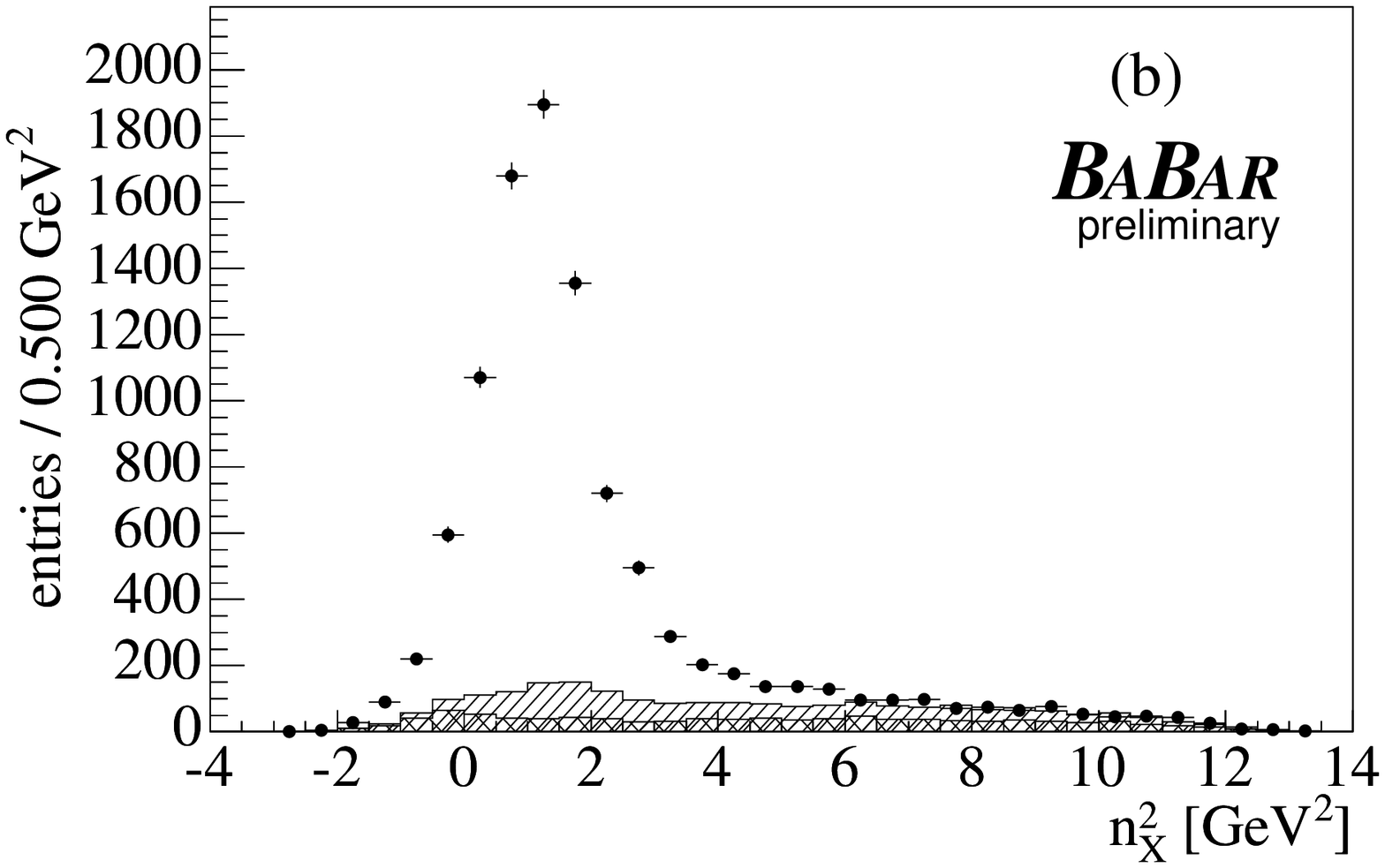}}%
        \caption{\it (a) The measured hadronic mass spectrum and (b) the 
measured $n_X$ spectrum for $E_l > 0.8\unit{GeV}$ in the $B$ rest frame. 
Tag-side backgrounds are
indicated by the hatched histogram, signal-side backgrounds by the 
cross-hatched histogram.}
\label{btocfig}
    \end{center}
\end{figure}

The measurement is based on a sample of $232$ million $B\bar{B}$ pairs.
After reconstructing the \Breco\ and identifying the lepton, 
where both electrons and muons are used, the hadronic system is reconstructed
from the remaining tracks and neutral energy depositions in the event. A
kinematic fit imposing energy-momentum conservation and the missing energy
and momentum in the event to be consistent with coming from one neutrino is 
performed to improve the resolution in the hadronic variables.
The hadronic mass and mixed hadronic mass-energy spectra are shown in 
Fig.~\ref{btocfig}.
The moments are extracted directly from the kinematically fitted hadronic
masses and energies and are corrected for the effect of lost particles.
The main contribution to the systematic uncertainties on the moments arise 
from the impact of the reconstruction efficiencies of neutral particles 
on the inclusive event reconstruction.

\begin{figure}[t]
    \begin{center}
        {\includegraphics[width=0.5\textwidth]{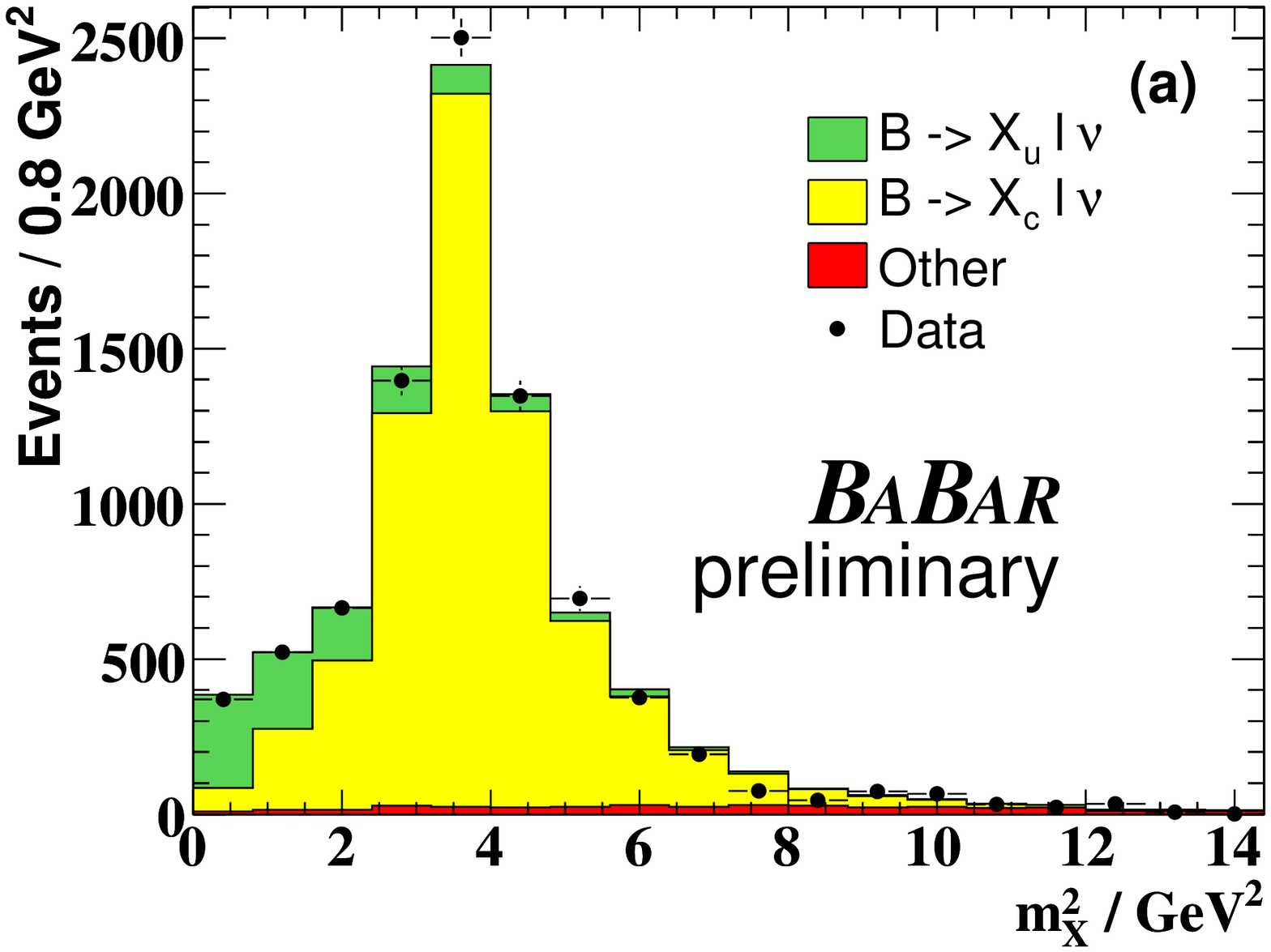}}\hfill
        {\includegraphics[width=0.5\textwidth]{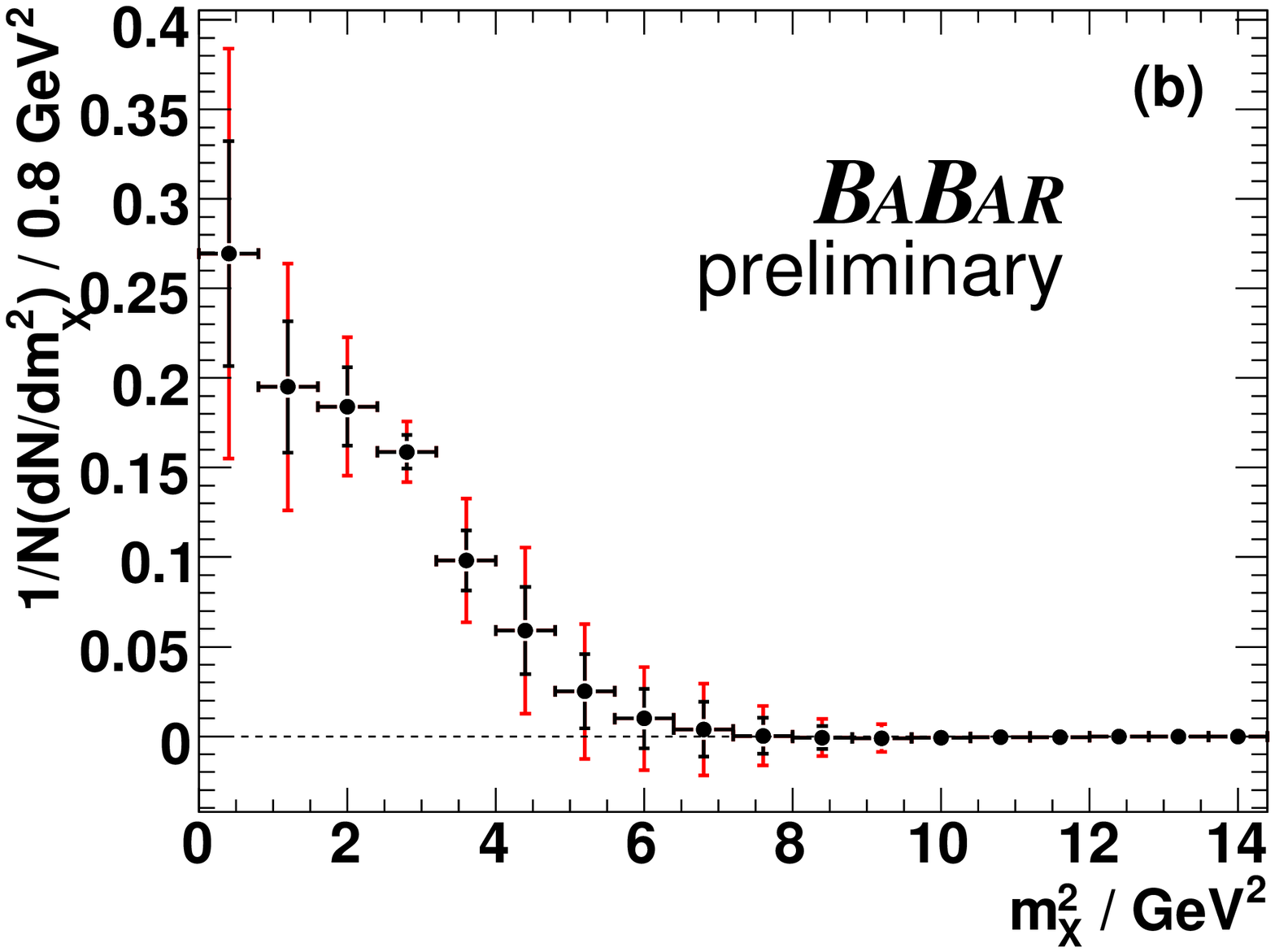}}%
        \caption{\it (a) The measured hadronic mass spectrum before subtraction
of \BtoXclnu\ and non-semileptonic backgrounds. (b) Unfolded hadronic mass spectrum
in \BtoXulnu. The inner error bars show the statistical uncertainties only.}
\label{btoufig}
    \end{center}
\end{figure}

A combined fit is performed to hadronic mass moments,  
lepton energy moments~\cite{Aubert:2004td} in \BtoXclnu\ decays and 
photon energy 
moments in \BtoXsgamma\ decays~\cite{Aubert:2005cua} and yields
$m_b = (4.552\pm 0.055)\unit{GeV}$, 
$m_c = (1.070\pm 0.085)\unit{GeV}$ 
(correlation $\rho_{m_b m_c} = 0.96$), 
$\mupisq = (0.471\pm 0.070)\unit{GeV}^2$, 
$\muGsq = (0.330\pm 0.060)\unit{GeV}^2$,
$\rhoDcube = (0.220\pm 0.047)\unit{GeV}^3$ and
$\rhoLScube = (-0.159\pm 0.095)\unit{GeV}^3$ in the kinetic scheme.

\section{Charmless Semileptonic $B$ Decays}
Moments of the hadronic mass spectrum in \BtoXulnu\ decays are used for a
preliminary extraction of \mb, \mupisq\ and \rhoDcube. Their determination in
\BtoXulnu\ allows for a test of
the theoretical framework used for the extraction of \Vub\ in the
same channel in which \Vub\ is determined. The hadronic mass moments are 
measured
with an upper cut on the hadronic mass to reduce experimental uncertainties. 

The measurement is based on a sample of $383$ million $B\bar{B}$ pairs.
After reconstructing the \Breco\ and identifying the lepton, 
where both electrons and muons with a minimum energy of $E_\ell = 1\unit{GeV}$
in the $B$ rest frame are used, the hadronic system is reconstructed
from the remaining tracks and neutral energy depositions in the event. Vetos
on identified $K^\pm$, reconstructed $K_S$ and partically reconstructed 
$D^{*\pm}$
are employed to suppress the dominant background from \BtoXclnu\ events. The
remaining \BtoXclnu\ and non-semileptonic backgrounds are subtracted by a fit
to the hadronic mass spectrum (Fig.~\ref{btoufig} (a)). 
The full \mxsq\ region contains $1027\pm 176$
signal events. The background-subtracted spectrum is unfolded for detector
acceptance, efficiency and resolution effects (Fig.~\ref{btoufig} (b)) 
and the first and second and 
third
central moments are extracted from the unfolded spectrum for 
$\mxsq < 6.4\unit{GeV}^2$:

\begin{equation*}
\begin{aligned}
\langle \mxsq \rangle 
&= (1.96 \pm 0.34 \mathrm{(stat)} \pm 0.53 \mathrm{(syst)})\unit{GeV}^2\\ 
\langle (\mxsq)^2 - \langle \mxsq\rangle^2 \rangle 
&= (1.92 \pm 0.59 \mathrm{(stat)} \pm 0.87 \mathrm{(syst)})\unit{GeV}^4\\
\langle (\mxsq)^3 - \langle \mxsq\rangle^3 \rangle 
&= (1.79 \pm 0.62 \mathrm{(stat)} \pm 0.78 \mathrm{(syst)})\unit{GeV}^6
\end{aligned}
\end{equation*}
with correlation coefficients $\rho_{12} = 0.99$, 
$\rho_{23} = 0.94$ and $\rho_{13} = 0.88$. The main systematic uncertainties
arise from the control of the \BtoXclnu\ background.

A fit of these moments to predictions in the kinetic 
scheme~\cite{Gambino:2005tp} yields 

\begin{equation*}
\begin{aligned}
m_b &= (4.604 \pm 0.125 \mathrm{(stat)} \pm 0.193 \mathrm{(syst)} \pm
0.097 \mathrm{(theo)})\unit{GeV}\\ \mupisq &= (0.398 \pm 0.135
\mathrm{(stat)} \pm 0.195 \mathrm{(syst)} \pm 0.036
\mathrm{(theo)})\unit{GeV}^2\\ \rhoDcube &= (0.102 \pm 0.017
\mathrm{(stat)} \pm 0.021 \mathrm{(syst)} \pm 0.066
\mathrm{(theo)})\unit{GeV}^3.
\end{aligned}
\end{equation*}
with correlation coefficients 
$\rho_{\mb\mupisq} = -0.99$, 
$\rho_{\mupisq\rhoDcube} = 0.57$ and $\rho_{\mb\rhoDcube} = -0.59$.

\section{Summary}

\begin{figure}[t]
    \begin{center}
      {\includegraphics[width=0.5\textwidth]{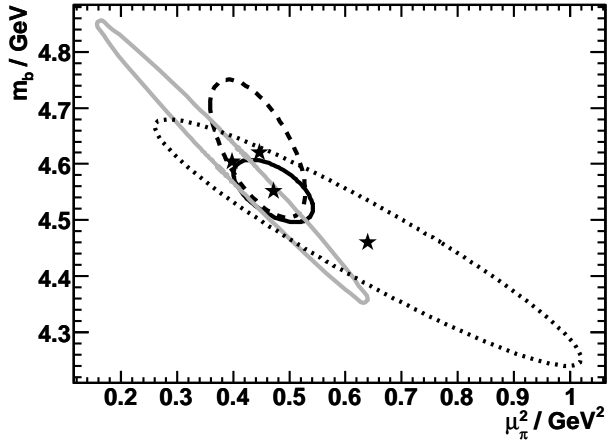}}
        \caption{\it Results from the recent \babar\ analyses presented in the 
\mb--\mupisq\ plane in the kinetic scheme. The dotted ellipse shows the result of
the presented \BtoXsgamma\ analysis~\cite{bsg}, the dashed ellipse the result
from \BtoXclnu\ decays~\cite{:2007yaa,Aubert:2004td}, 
the black solid ellipse the result
from its combination with earlier \BtoXsgamma\ 
measurements~\cite{Aubert:2005cua} and the solid grey ellipse the result from
\BtoXulnu\ decays.}
\label{mbmupisqfig}
    \end{center}
\end{figure}
We presented preliminary determinations of the $b$-quark mass \mb\ and 
nonperturbative
parameters from charmed and charmless semileptonic and radiative penguin
$B$ decays at \babar. The determination in charmless 
semileptonic $B$ decays has been performed for the first time.
The results for \mb\ and \mupisq\ are summarized in 
Table~\ref{resultstable} and compared in Fig.~\ref{mbmupisqfig}.
The determinations in the different channels are consistent within the quoted
uncertainties and with earlier 
determinations~\cite{Buchmuller:2005zv}.
\begin{table}[t]
\centering
\caption{ \it Results for \mb\ and \mupisq\ in the kinetic scheme and their correlations. 
}
\vskip 0.1 in
\begin{tabular}{|l|c|c|c|} \hline
          &  \mb$/\unit{GeV}$ & \mupisq$/\unit{GeV}^2$ & $\rho$\\
\hline
\hline
 \BtoXsgamma~\cite{bsg}   & $4.46^{+0.21}_{-0.23}$   & 
                           $0.64^{+0.39}_{-0.38}$   & $-0.94$            \\
 \BtoXclnu~\cite{:2007yaa,Aubert:2004td} and \BtoXsgamma~\cite{Aubert:2005cua}
                          & $4.552\pm 0.055$  &
                           $0.471\pm 0.070$  & $-0.56$\\

 \BtoXulnu               & $4.604\pm 0.250$  &
                           $0.398\pm 0.240$  & $-0.99$\\
\hline
\end{tabular}
\label{resultstable}
\end{table}

Work supported in part by the Department of Energy contract
DE-AC02-76SF00515.

\end{document}